\documentclass{article}  
\usepackage{spadre2008}
\usepackage{graphicx}
\frompage{000} \topage{000}                                              
\def\rightmark{Integrated Boltzmann+Hydrodynamics}
\def\leftmark{H. Petersen et al.}
 
\title{A fully integrated Boltzmann+hydrodynamics approach: Multiplicities, transverse dynamics and HBT} 
\authors{
{Hannah Petersen$^{1,2}$, Jan Steinheimer$^{2}$, Qingfeng Li$^{1}$, Gerhard Burau$^{2}$ and Marcus Bleicher$^{2}$ %
}\\[2.812mm]
{\normalsize
\hspace*{-8pt}$^1$ Frankfurt Institute for Advanced Studies (FIAS),\\ 
Ruth-Moufang-Str.~1, D-60438 Frankfurt am Main, Germany\\[0.2ex] 
\hspace*{-8pt}$^2$ Institut f\"ur Theoretische Physik, Goethe-Universit\"at,\\ 
Max-von-Laue-Str.~1, D-60438 Frankfurt am Main, Germany
}}
 
\abstract{We present a coupled Boltzmann and hydrodynamics approach to relativistic heavy ion reactions. This hybrid approach is based on the Ultra-relativistic Quantum Molecular Dynamics (UrQMD) transport approach with an intermediate hydrodynamical evolution for the hot and dense stage of the collision. This implementation allows to compare microscopic transport calculations with hydrodynamic calculations using exactly the same initial conditions and freeze-out procedure. Here, the results of the different calculations for particle multiplicities, the mean transverse mass excitation function and the HBT radii at all SPS energies are discussed in the context of the available data.}

 
\begin{document}
 
\maketitle
\setcounter{page}{1}

One of the main motivations to study high energy heavy ion collisions is the creation of a new deconfined phase of strongly interacting matter, the so called Quark-Gluon Plasma (QGP) \cite{Bass:1998vz}. Since the direct detection of free quarks and gluons is impossible due to the confining nature of QCD, it is important to model the dynamical evolution of heavy ion reactions to draw conclusions from the final state particle distributions about the interesting early stage of the reaction. To get a more consistent picture of the whole dynamics of heavy ion reactions various so called microscopic plus macroscopic (micro+macro) hybrid approaches have been launched during the last decade \cite{Dumitru:1999sf,Bass:1999tu,Bass:2000ib,Soff:2000eh,Soff:2001hc,Nonaka:2006yn,Andrade:2006yh,Steinheimer:2007iy}). 

In this paper we briefly describe the specific micro+macro hybrid approach that embeds a hydrodynamic phase in the UrQMD approach. This allows to reduce the parameters for the initial conditions and the freeze-out prescription. At present calculations imposing a hadron gas EoS are shown to provide a baseline to disentangle the effects of the different assumptions for the underlying dynamics in a transport vs. hydrodynamic calculation. We show results for particle multiplicity, HBT radii and $\langle m_T \rangle$ excitation functions in the context of the available data at FAIR/SPS energies.

The Ultra-relativistic Quantum Molecular Dynamics Model (UrQMD) is used to calculate the initial state of a heavy ion collision for the hydrodynamical evolution \cite{Steinheimer:2007iy,Bass:1998ca,Bleicher:1999xi}. This has been done to account for the non-equilibrium dynamics in the very early stage of the collision. In this configuration the effect of event-by-event fluctuations of the initial state are naturally included. The coupling between the UrQMD initial state and the hydrodynamical evolution proceeds when the two Lorentz-contracted nuclei have passed through each other.

\begin{figure}[htb]
\vspace*{-.8cm}
\centering
\includegraphics[angle=0,width=0.7\textwidth]{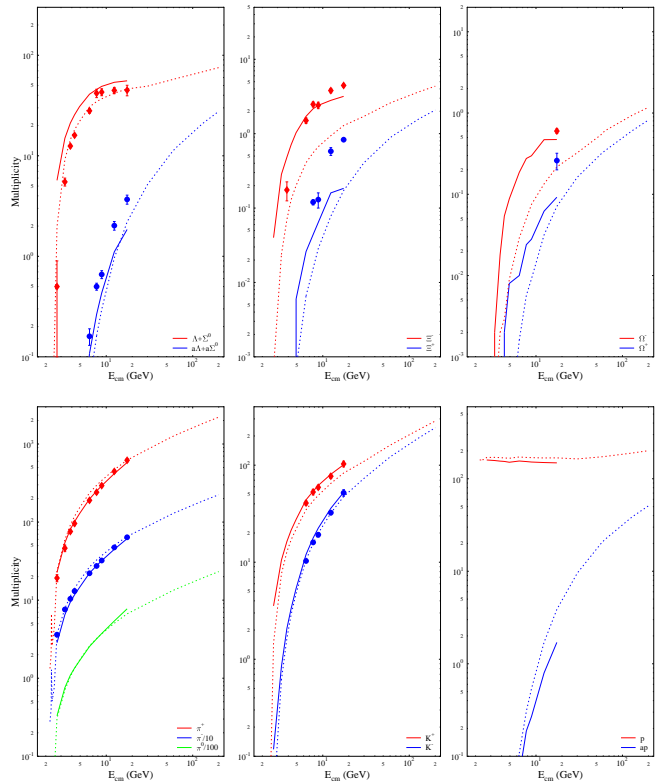}
\vspace*{-1cm}
\caption[]{(Color online) Excitation function of particle multiplicities ($4\pi$) in Au+Au/Pb+Pb collisions from $E_{\rm lab}=2A~$GeV to $\sqrt{s_{NN}}=200$ GeV. UrQMD+Hydro (HG) calculations are depicted with full lines, while UrQMD-2.3 calculations are depicted with dotted lines. The corresponding data from different experiments \cite{Klay:2003zf,Pinkenburg:2001fj,Chung:2003zr,:2007fe,Afanasiev:2002mx,Anticic:2003ux,Richard:2005rx,Mitrovski:2006js,arXiv:0804.3770,Blume:2004ci,Afanasiev:2002he,Alt:2004kq} are depicted with symbols.}
\label{fig_mulally}
\end{figure}

After the UrQMD initial stage, a full (3+1) dimensional hydrodynamic evolution is performed using the SHASTA algorithm \cite{Rischke:1995ir,Rischke:1995mt}. For the results presented here an equation of state for a free hadron gas without any phase transition is used \cite{Zschiesche:2002zr}.

The hydrodynamic evolution is stopped, if the energy density drops below five times the ground state energy density (i.e. $\sim 730 {\rm MeV/fm}^3$) in all cells. This criterium corresponds to a T-$\mu_B$-configuration where the phase transition is expected - approximately $T=170$ MeV at $\mu_B=0$. The hydrodynamic fields are mapped to particle degrees of freedom via the Cooper-Frye equation on an isochronous hypersurface. The particle vector information is then transferred back to the UrQMD model, where rescatterings and the final decays are perfomed using the hadronic cascade. A more detailed description of the hybrid model including parameter tests and results for multiplicities and spectra can be found in \cite{Petersen:2008dd}. 

In the following results from the UrQMD+Hydro (HG) model are compared to the default transport calculation (UrQMD-2.3) and to the experimental data. Note that parameters are constant for different energies and/or centralities in the hybrid model calculation.

\begin{figure}[htb]
\vspace*{-.8cm}
\centering
\includegraphics[width=0.6\textwidth]{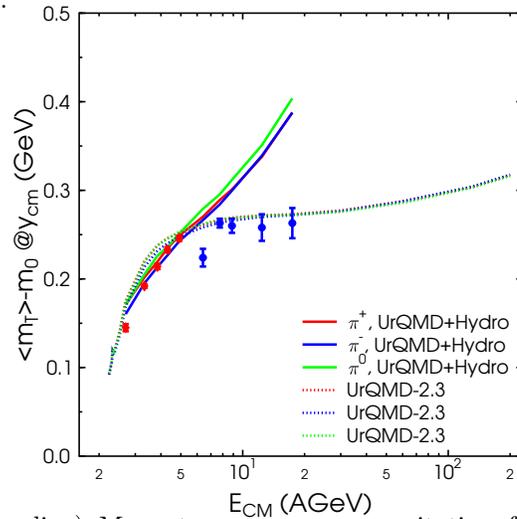}
\vspace*{-1cm}
\caption[]{(Color online) Mean transverse mass excitation function of pions at midrapidity ($|y|<0.5$) for central ($b<3.4$ fm) Au+Au/Pb+Pb collisions from $E_{\rm lab}=2-160A~$GeV. UrQMD+Hydro (HG) calculations are depicted with full lines, while UrQMD-2.3 calculations are depicted with dotted lines. The corresponding data from different experiments \cite{Ahle:1999uy,:2007fe,Afanasiev:2002mx} are depicted with symbols.}
\label{fig_mmt_exc}
\end{figure} 

Fig. \ref{fig_mulally} shows the excitation functions of the total multiplicities for central Au+Au/Pb+Pb collisions for $E_{\rm lab}=2A~$GeV to $\sqrt{s_{\rm NN}}=200$ GeV. Compared to the default simulation, the pion and proton multiplicities are decreased over the whole energy range in the hybrid model calculation due to the conservation of entropy in the ideal hydrodynamic evolution. The non-equilibrium transport calculation produces entropy and therefore the yields of pions are higher. The production of strange particles however, is enhanced due to the establishment of full local equilibrium in the hybrid calculation. Since the yield of strange particles is small they survive the interactions in the UrQMD evolution that follows the hydrodynamic freeze-out almost without re-thermalization.
    
Next, we turn to the excitation function for the mean transverse mass of pions. The UrQMD approach shows a softening of the equation of state in the region where the phase transition is expected because of excited resonances and non-equilibrium dynamics while the hydrodynamic calculation with hadron gas EoS just rises as a function of the energy.  

\begin{figure}[htb]
\vspace*{-.8cm}
\centering
\includegraphics[angle=0,width=\textwidth]{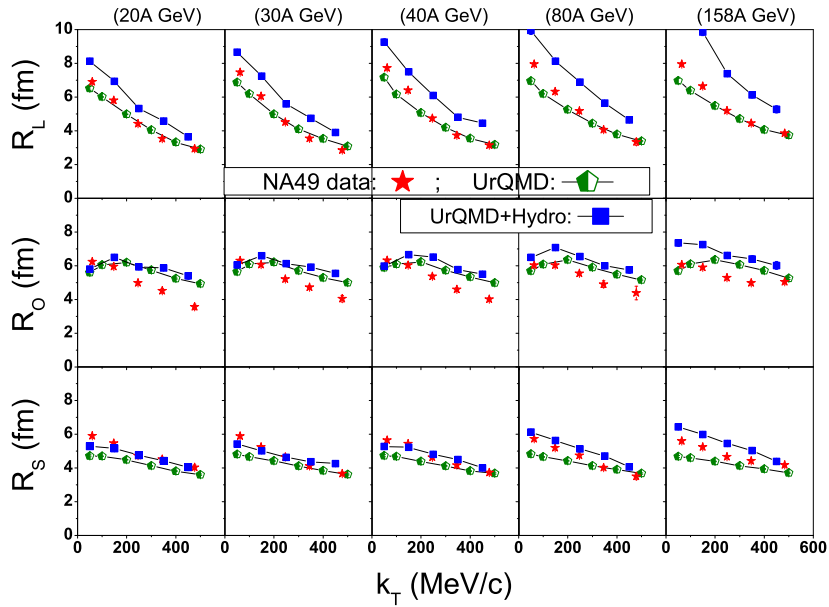}
\vspace*{-1cm}
\caption[]{(Color Online)$k_T$ dependence of the
Pratt-radii of the pion source in central Pb+Pb collisions at
$E_{\rm lab}=20$, $30$, $40$, $80$, and $158$A GeV and mid-rapidity
$0<Y_{\pi\pi}<0.5$. Preliminary NA49 data are taken from
\cite{Kniege:2006in}.}
\label{fig_hbt}
\end{figure}

Finally we show HBT results. It is well known that the Hanbury-Brown-Twiss interferometry (HBT)
technique can provide important information about the
spatio-temporal structure of the particle emission source (the
region of homogeneity) \cite{Rischke:1996em}. The HBT radii can be extracted by
fitting the correlator of the particle pair with a Gaussian form. Using the Pratt's three-dimensional convention (the
LCMS system), the parametrization of the correlation function reads $C(q_L,q_O,q_S)=1+\lambda e^{-R_L^2q_L^2-R_O^2q_O^2-R_S^2q_S^2-2R_{OL}^2q_Oq_L}$ where $\lambda$ is normally referred to as an incoherence factor,
but we can regard it as a free parameter since it might be affected
by many other factors. $R_L$, $R_O$, and $R_S$ are the Pratt radii
in longitudinal, outward, and sideward directions, while the
cross-term $R_{OL}$ plays a role at large rapidities. $q_i$ is the
pair relative momentum $\mathbf{q}=\mathbf{p}_1-\mathbf{p}_2$ in the $i$ direction.

Fig. \ref{fig_hbt} shows the transverse momentum $k_T$ dependence
($\textbf{k}_T=(\textbf{p}_{1T}+\textbf{p}_{2T})/2$) of the HBT-
radii $R_L$ (top plots), $R_O$ (middle plots), and $R_S$ (bottom
plots) of the negatively charged pion source in Pb+Pb reactions at
beam energies $E_{\rm lab}=20$, $30$, $40$, $80$, and $158$A GeV. The
calculations (lines with symbols) are compared with preliminary NA49
data (stars) \cite{Kniege:2006in} ($<7.2\%$ of the total cross
section $\sigma_T$). The pair-rapidity $0<Y_{\pi\pi}<0.5$ is chosen for all reactions. 

One observes that the calculations in the UrQMD cascade mode can reproduce the
$k_T$-dependence of HBT radii $R_L$ and $R_S$ fairly well. Only at
small $k_T$ values, the calculated $R_L$ and $R_S$ values are up to $25\%$ lower than data \cite{Li:2007yd}. For the $R_O$ values, the calculations reach higher values than the experimental data especially at relatively
large $k_T$. The HBT $R_L$ radii from the hybrid model are seen to be larger than data and those from cascade UrQMD
calculations. Interestingly, however is the fact that the stronger expansion leads to a better description of the $R_s$ radii, improving the $R_o/R_s$ ratios. Certainly,
the EoS used in hydro-process in the hybrid model will also strongly influence
the final HBT radii and will be investigated in a forthcoming paper.

\begin{figure}[h]
\vspace*{-.8cm}
\centering
\includegraphics[width=0.8\textwidth]{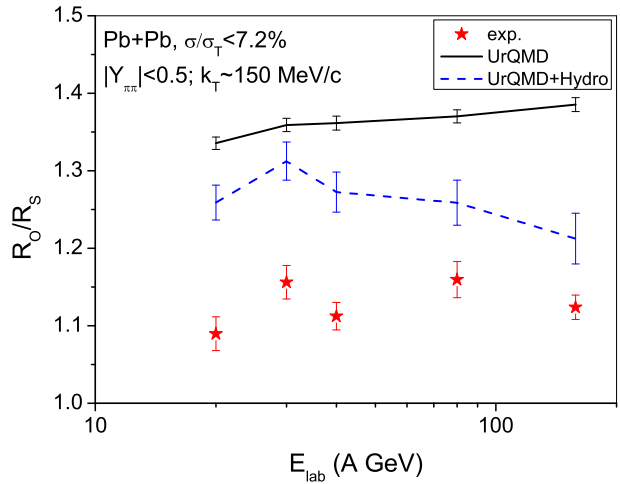}
\vspace*{-.5cm}
\caption[]{(Color online) Beam-energy dependence of the ratio $R_O/R_S$ at $100<k_T<
200$ MeV$/c$ for central Pb+Pb collisions at SPS energies.}
\label{fig_rorsexc}
\end{figure}  

Fig. \ref{fig_rorsexc} illustrates the beam-energy dependence of the ratio
$R_O/R_S$ of the pion source at $100<k_T< 200$ MeV$/c$ for central
Pb+Pb collisions at SPS energies. In the cascade mode, the
calculated ratio is higher than the experimental data, while with
the embedded hydrodynamic evolution the ratio is driven down towards the experimantal data. With
increasing beam energy, the difference of the $R_O/R_S$ ratio
between the default transport and the hybrid model calculation becomes larger. 

We have presented the main ideas of an integrated Boltzmann and hydrodynamics approach to relativistic heavy ion reactions. The final pion and proton multiplicities are lower in the hybrid model calculation due to the isentropic hydrodynamic expansion while the yields for strange particles are enhanced due to the local equilibrium in the hydrodynamic evolution. The results of the different calculations for the mean transverse mass excitation function and the HBT radii at all FAIR/SPS energies are sensitive to the underlying dynamics and reflect the stronger transverse expansion in the hybrid model.

\section*{Acknowledgments}
We are grateful to the Center for the Scientific Computing (CSC) at Frankfurt for the computing resources. The authors thank Dirk Rischke for providing the 1 fluid hydrodynamics code. H. Petersen gratefully acknowledges financial support by the Deutsche Telekom-Stiftung and support from the Helmholtz Research School on Quark Matter Studies. This work was supported by GSI and BMBF. 
 

\vfill\eject
\end{document}